\documentclass[pre,preprint,showpacs,superscriptaddress]{revtex4}

\begin{document}
\bibliographystyle{apsrev}
\title{ 
Exact Results
for Hydrogen Recombination
on Dust Grain Surfaces
}
\author{
Ofer Biham}  
\affiliation{
Racah Institute of Physics, 
The Hebrew University, 
Jerusalem 91904, Israel 
}
\author{
Azi Lipshtat}  
\affiliation{
Racah Institute of Physics, 
The Hebrew University, 
Jerusalem 91904, Israel 
}

\begin{abstract}

The recombination of hydrogen 
in the interstellar medium, 
taking place on surfaces of microscopic
dust grains, 
is an essential process in the evolution of chemical complexity
in interstellar clouds. 
Molecular hydrogen plays an important role in absorbing the heat
that emerges
during gravitational collapse, thus enabling the formation of structure
in the universe.
The H$_2$ formation process has been studied theoretically, and
in recent years also by laboratory experiments.
The experimental results were analyzed using a rate equation model.
The parameters of the surface, 
that are relevant to H$_2$ formation,
were obtained and used in order to calculate the recombination rate
under interstellar conditions.
However, it turned out that
due to the microscopic size of the dust grains and the low density
of H atoms, the rate equations may not always apply.
A master equation approach that provides a good
description of the H$_2$ formation process was proposed.
It takes
into account both the discrete nature of the H
atoms and the fluctuations in the number of atoms on a grain. 
In this paper we present 
a comprehensive analysis of the H$_2$ formation process,
under steady state conditions,
using
an exact solution of the master equation.
This solution provides 
an exact result for the
hydrogen recombination rate 
and its dependence on the flux, the surface temperature and the grain size.
The results are compared with those obtained from the rate equations.
The relevant length scales in the problem are identified and the parameter
space is divided into two domains.
One domain, characterized by first order kinetics, 
exhibits high efficiency
of H$_2$ formation.
In the other domain, characterized by second order kinetics,
the efficiency of H$_2$ formation is low.
In each of these domains we identify the range of parameters in which, due
to the small size of the grains, the rate equations do not 
account correctly for the recombination rate
and the
master equation is needed.

\end{abstract}


\maketitle


\section{Introduction}

The recombination of hydrogen on the surfaces of microscopic dust grains in the
interstellar medium has attracted much interest in recent years.
This process is essential since
gas-phase reactions cannot account for the abundance of
H$_2$ in interstellar clouds
\cite{Gould1963,Hollenbach1970,Hollenbach1971a,Hollenbach1971b}. 
Theoretical
\cite{Williams1968,Smoluchowski1981,Smoluchowski1983,Aronowitz1985,Duley1986,Pirronello1988,Sandford1993,Takahashi1999}  
and
experimental 
\cite{Brackmann1961,Schutte1976,Pirronello1997a,Pirronello1997b,Pirronello1999,Manico2001}
techniques have been used in order to 
evaluate the rate of H$_2$ formation on relevant dust materials 
under interstellar conditions.
Quantum mechanical calculations
and molecular dynamics simulations were performed, in attempt to identify the
diffusion and reaction rates on the surfaces of various 
astrophysically
relevant materials
\cite{Hollenbach1970,Hollenbach1971a,Hollenbach1971b,Smoluchowski1981,Smoluchowski1983,Buch1991a,Buch1991b,Takahashi1999}.
Experimental 
results were also obtained for
the activation
energies of the relevant diffusion and desorption processes on
various surfaces
\cite{Pirronello1997a,Pirronello1997b,Pirronello1999}.

Rate equations are an essential tool in the modeling of chemical
reactions in the interstellar medium
\cite{Pickles1977,Hendecourt1985,Brown1990a,Brown1990b,Hasegawa1992,Hasegawa1993a,Hasegawa1993b,Caselli1993,Caselli1994,Ruffle2000,Ruffle2001a,Ruffle2001b}.
Chemical models based on the rate equation approach
take into account a large number of reactions
in the gas phase as well as reactions that take place on the
surfaces of dust grains.
Rate equations were recently used 
in order to analyze
the results of laboratory experiments on H$_2$ formation on
dust-analog surfaces
\cite{Katz1999}.
The analysis provided the surface parameters that are essential
for the evaluation of the H$_2$ formation rate on dust grains 
in the interstellar medium.
The rate equations used in Ref. 
\cite{Katz1999}
describe the diffusion, reaction and desorption processes on the surface.
They provide
the time evolution of the average densities of atoms and molecules
on the surface, while fluctuations are neglected.
Such rate equations
are expected to
provide good results for macroscopic surfaces. 
However, it turns out that 
they may be unsuitable for the study
of hydrogen recombination in the
interstellar medium
due to the small grain size 
and low flux
\cite{Tielens1995,Charnley1997,Caselli1998,Shalabiea1998,Stantcheva2001}. 
Under these conditions the number of H atoms on the surface of 
a grain may be very small and fluctuations 
are expected to be
significant.
Attempts to resolve this difficulty included 
the use of modified rate equations in which parameters
are changed to account for the finite grain size 
\cite{Tielens1995,Charnley1997,Caselli1998,Shalabiea1998,Stantcheva2001}. 
Monte Carlo methods were also used, to simulate the surface diffusion and reaction 
processes on small grains
\cite{Charnley1998,Charnley2001}.
The class of Monte Carlo methods that are suitable for such activated processes
on the surface, away from thermal equilibrium, are the continuous time or kinetic
Monte Carlo techniques
\cite{Barkema1999}.
In these simulations, at each time step, the next move is picked with a probability
proportional to its rate.
The elapsed time is given, according to the theory of stochastic processes,
by the inverse of the sum of the rates of all processes that could have
occurred at that time.
The kinetic Monte Carlo approach can be directly related to the underlying
master equation that describes the time evolution of the probabilities
of all the microscopic states of the system
\cite{vanKampen1981}.
Monte Carlo simulations typically require large computational resources.
For example, the calculation of averages for quantities such as the H$_2$
formation rate 
is done by collecting
large amounts of statistical information 
\cite{Charnley1998,Charnley2001}.
To study the chemistry of interstellar clouds one needs a model that  
couples the gas-phase and grain-surface reactions.  
It was found that the use of Monte Carlo methods in this context
is impractical, while rate equations for surface chemistry cannot
account correctly for reaction rates on small grains.

Recently it was shown that the H$_2$ formation process on small
grains can be described by a master equation approach
\cite{Biham2001,Green2001}.
The master equation 
takes into account both the discrete nature of the
H atoms as well as the fluctuations. 
Its dynamical variables are the
probabilities $P (n)$ 
that there are 
$n$ 
atoms
on a grain
at time $t$. 
The time derivatives  
$\dot{P} (n )$, $n =0, 1, 2, \dots$
are expressed 
in terms of the adsorption, reaction and desorption terms.
The master equation provides the time evolution of 
the probabilities
$P (n )$, 
from which
the recombination rate can be calculated.
It was used
in conjunction with surface parameters obtained experimentally
and computationally,
to explore the hydrogen recombination
and other chemical reactions 
on small grains under interstellar conditions
\cite{Biham2001,Green2001,Stantcheva2002}.

In this paper we present a comprehensive analysis of the 
H$_2$ formation process 
under steady state conditions as a function of the physical
parameters.
These parameters include the flux of H atoms, the grain size and surface
temperature.
They also include properties of the surface, namely the
density of adsorption sites 
as well as the activation energies for H diffusion and desorption.
The analysis is based on an exact analytical solution of the master
equation.
In this solution, 
the steady state distribution  $P(n)$
is expressed in terms of two dimensionless quantities, 
composed of the physical parameters mentioned above.
Using this solution we identify the
relevant length scales in the problem.
The parameter
space is then divided into two domains.
One domain, characterized by first order kinetics 
(namely, the H$_2$ formation rate is linearly proportional to the
flux of H atoms), 
exhibits high efficiency
of H$_2$ formation.
In the other domain, characterized by second order kinetics
(namely, the H$_2$ formation rate is proportional to the square
of the incoming flux),
the efficiency of H$_2$ formation is low.
In each of these domains we identify the range of parameters in which, due
to the small size of the grains, the rate equations do not apply and the
master equation is needed.

The paper is organized as follows. 
The rate equation model is 
described in Sec. IIA and analyzed in Sec. IIB.
The master equation is introduced in Sec. IIIA. 
The exact solution of the master equation is
given in Sec. IIIB.
The analysis of the H$_2$ formation process
on small grains, 
that is based on this solution,
is presented in Sec. IIIC.
The results are discussed in Sec. IV and summarized in Sec. V.

\section{H$_2$ Formation on Macroscopic Surfaces}

\subsection{The Rate Equation Model}
\label{sec:RateModel}

Consider a flux of H atoms 
that are irradiated 
and stick to a macroscopic surface.
The atoms perform hops as random walkers on the surface 
and recombine into H$_2$ molecules when they encounter one another.
Let $\rho(t)$ (in monolayers [ML]) be the coverage 
of H atoms on the surface
at time $t$. 
Its time dependence is described by
the following rate equation:

\begin{equation}
{d \rho \over dt}  =  
f
- W \rho - 2 a \rho^2. 
\label{eq:N}
\end{equation}

\noindent
The first term on the right hand side of 
Eq.~(\ref{eq:N}) 
represents the 
flux of H atoms.
The parameter
$f$ represents the 
{\em effective} flux
of atoms 
(in units of ML $s^{-1}$), 
namely,
the 
(temperature dependent)
sticking coefficient
$\xi(T)$
is absorbed into 
$f$
[the Langmuir-Hinshelwood (LH) rejection process 
\cite{Zangwill1988}
of atoms deposited on top
of already adsorbed atoms is neglected here since it is 
assumed that the coverage is low].
The second term in 
Eq.~(\ref{eq:N}) 
represents the desorption of H atoms from the
surface. 
The desorption coefficient is 

\begin{equation}
W =  \nu \cdot \exp (- E_{1} / k_{B} T)  
\label{eq:P1}
\end{equation}

\noindent
where $\nu$ is the attempt rate 
(standardly taken to be $10^{12}$ s$^{-1}$), 
$E_{1}$ 
is the activation energy barrier for desorption 
of an H atom and $T$ is the temperature.
The third term in 
Eq.~(\ref{eq:N}) 
accounts for the depletion of the H population
on the surface due to recombination into H$_{2}$ molecules, 
where

\begin{equation} 
a =  \nu \cdot \exp (- E_{0} / k_{B} T) 
\label{eq:Alpha}
\end{equation}

\noindent
is the hopping rate of H atoms on the surface
and $E_{0}$ is the activation energy barrier for hopping.
Here we assume that diffusion occurs only by thermal hopping,
in agreement with experimental results
\cite{Katz1999}.
We also assume that there is no energy
barrier for recombination. 
The H$_{2}$ production rate $r$
(ML s$^{-1}$) 
is given by  

\begin{equation}
r  =   a \ \rho^{2}. 
\label{eq:Production}
\end{equation}

\noindent
For simplicity 
we assume here 
that all the H$_2$ molecules are desorbed from the surface
upon formation. Even if on real surfaces some of the molecules
may
remain on the surface and desorb thermally later, 
under steady state conditions,
at low coverage, this will not affect the recombination rate.

A (more complete) model, based on
Eq.~(\ref{eq:N})
was used 
\cite{Katz1999}
to analyze the results
of temperature programmed desorption (TPD) experiments 
\cite{Pirronello1997a,Pirronello1997b,Pirronello1999}. 
The values of
$E_0$ 
and
$E_1$ 
(as well as two additional parameters)
that best fit the experimental results were obtained.
The steady state behavior under astrophysically relevant conditions
was then studied and the
recombination efficiency 

\begin{equation}
\eta = {r \over {f/2}}
\end{equation}

\noindent
was calculated in an astrophysically relevant range of flux and temperature.
Note that under steady state conditions $\eta$
is limited to the range 
$0 \le \eta \le 1$.
It  was found that the
recombination efficiency is highly temperature dependent. 
It exhibits a narrow window of high efficiency along the temperature axis,
which slowly shifts to higher temperatures as the flux is increased
\cite{Katz1999}. 

\subsection{Analysis and Results}

Consider a macroscopic surface exposed to a constant flux
of H atoms, as described by 
Eq.~(\ref{eq:N}).
Under steady state conditions 
$d \rho / dt = 0$
and the coverage is given by

\begin{equation}
\rho = {1 \over 4} \left( {W \over a} \right)
\left[ -1 + \sqrt{1 + 8 {(f/W) \over (W/a)} }   \right].
\label{eq:steadyrho}
\end{equation}

\noindent
The recombination efficiency $\eta=2(a/f)\rho^2$
takes the form

\begin{equation}
\eta = {1 \over 8} 
{(W/a) \over (f/W)}
\left[ -1 + \sqrt{1 + 8 {(f/W) \over (W/a)} }   \right]^2.
\label{eq:steadyeta}
\end{equation}

\noindent
To obtain a better understanding of the recombination process
we will try to identify
the length and time scales involved.
Consider the ratio $a/W$ between the hopping rate and the desorption rate.
This ratio is 
(on average) the number of hops an H atom visits before it
desorbs (neglecting recombination).
It is also
(up to a constant multiplicative factor 
of order unity and a logarithmic correction)
the number of sites 
that the atom visits
before it desorbs
\cite{Montroll1965}. 
We will denote this number by

\begin{equation}
s_{\rm visit} = a/W.
\end{equation}

\noindent
Consider the ratio $W/f$ between the desorption rate and the flux.
Neglecting the recombination term we obtain
$W/f = 1/\rho$,
namely this is approximately the average number of vacant sites around
each H atom.
We denote it by

\begin{equation}
s_{\rm vacant} = W/f.
\end{equation}

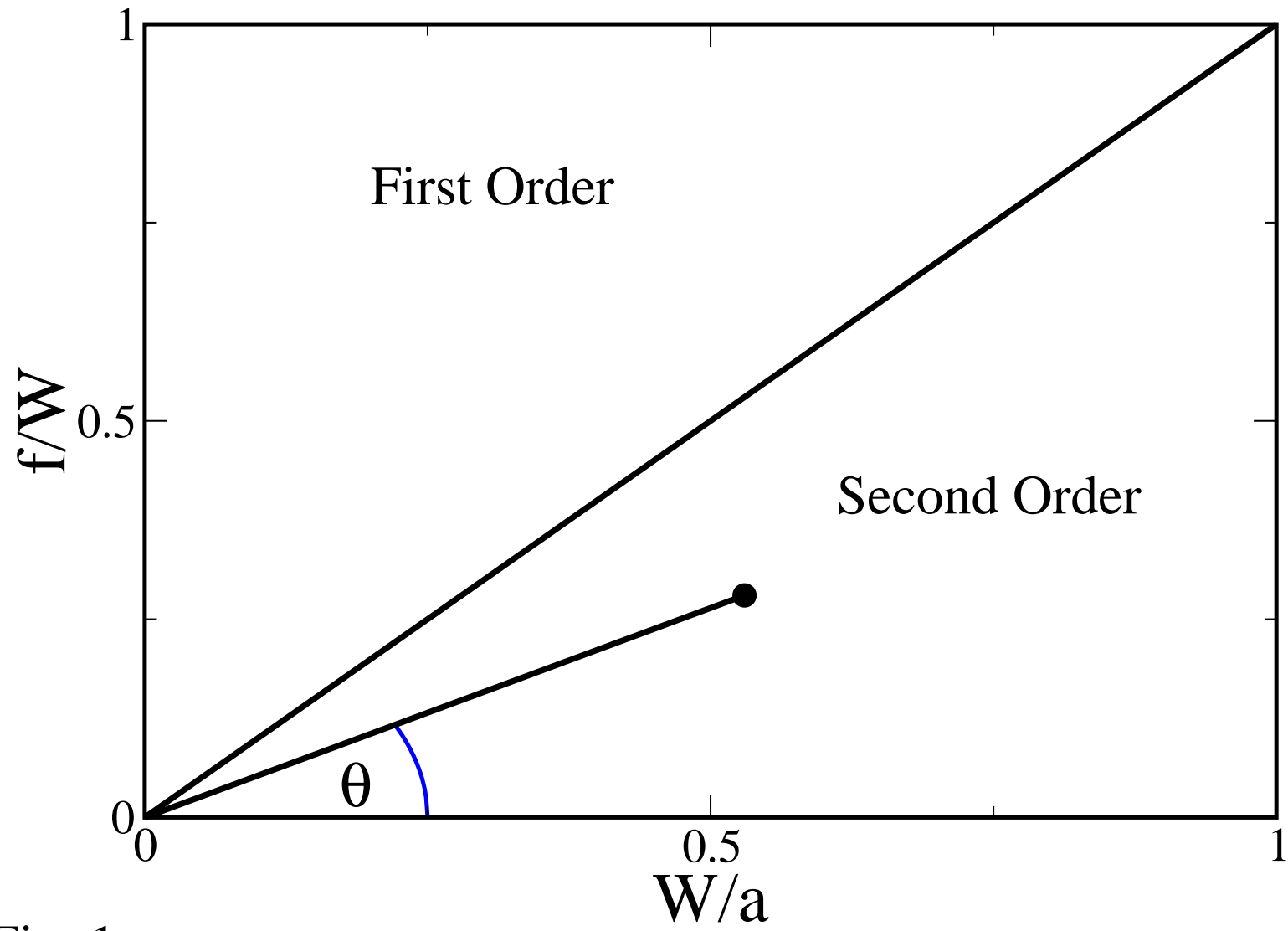
\begin{figure}
\caption{
(a) The phase diagram of hydrogen recombination on macroscopic
surfaces 
in the $(W/a,f/W)$ plane,
as described by Eq.~(\ref{eq:N}).
The parameter space is divided into two domains:
the first order domain, above the diagonal and 
the second order domain below it;
}
\label{fig:1}
\end{figure}

\noindent
The properties of the recombination process can thus be plotted
on a two dimensional parameter space, namely $f/W$ vs. $W/a$
[Fig. \ref{fig:1}].
The coverage $\rho$ depends on these two parameters, while $\eta$
depends only on the ratio 

\begin{equation}
\tan \theta = {(f/W) \over (W/a)}, \ \ \  0 \le \theta \le \pi/2.
\label{eq:theta}
\end{equation}

\noindent
The coverage $\rho$ 
and the recombination efficiency $\eta$
as a function of $\theta$ 
along the curve

\begin{equation}
{f \over W} = c \left( {W \over a} \right)^{-1},
\label{eq:hyperbole}
\end{equation}

\noindent
where $c=1/2$, are 
shown in 
Figs. \ref{fig:2}(a) 
and \ref{fig:2}(b), 
respectively.
The solid lines in 
Figs. 
\ref{fig:2}(a) 
and 
\ref{fig:2}(b)
show the results without LH rejection,
while the dashed lines show the results for
similar conditions but including LH rejection. 
In these Figures we identify a domain of 
high coverage and high efficiency for
$f/W > W/a$
and a domain of low coverage and low efficiency
for 
$f/W < W/a$,
separated by the diagonal line in Fig. \ref{fig:1}.
We will now analyze the limits, 
deep in each of these domains,
in terms of the length scales
associated with
$s_{\rm visit}$
and
$s_{\rm vacant}$
(the length-scales are obtained by dividing
$s_{\rm visit}$
and
$s_{\rm vacant}$
by the
density of adsorption sites and taking the square root).
In the case
$f/W \gg W/a$,
namely
$s_{\rm visit}
\gg
s_{\rm vacant}$,
the typical number of sites that an atom visits is much larger
than the number of vacant sites around it.
Therefore, it is most likely to find a second H atom and recombine.
In the opposite limit
$f/W \ll W/a$,
namely
$s_{\rm visit}
\ll
s_{\rm vacant}$,
most atoms visit only vacant sites around
their initial  
adsorption site
and desorb before having
a chance to form molecules.
As a result, the recombination efficiency $\eta$ is low.
Evaluating the coverage in both limits we obtain
\cite{Biham1998}

\begin{equation}
\rho \cong 
\left\{
\begin{array}{ll}
{1 \over \sqrt{2}} \sqrt{f \over a}:  & {f \over W} \gg {W \over a}         \\
{f \over W}: & {f \over W} \ll {W \over a}.
\end{array}
\right.
\label{limitsrho}
\end{equation}

\noindent
The H$_2$ production rate 
$r=a \rho2$
is given by

\begin{equation}
r \cong 
\left\{
\begin{array}{ll}
{1 \over 2} f:               & {f \over W} \gg {W \over a}         \\
{a \over W^2} f^2:           & {f \over W} \ll {W \over a}.
\end{array}
\right.
\label{limitsr}
\end{equation}

\noindent
In the first limit, $r$ is linear in the flux $f$, thus we denote this limit
as first order kinetics.
In the second case $r$ is proportional to $f2$
and we thus denote it as second order kinetics.
The regime of first order kinetics is characterized by high recombination efficiency
$\eta \cong 1$, since the desorption of H atoms is negligible.
In the second order kinetics, desorption of H atoms (before they form molecules) 
is the dominant process.
Therefore, the efficiency $\eta$ is low.
More precisely, the recombination efficiency is

\begin{equation}
\eta \cong 
\left\{
\begin{array}{ll}
1:               & {f \over W} \gg {W \over a}         \\
{2 a \over W^2} f:           & {f \over W} \ll {W \over a}.
\end{array}
\right.
\label{limitseta}
\end{equation}

For a given surface it is convenient to plot
the coverage and recombination rate as a function of the flux and the
surface temperature
\cite{Katz1999}.
The boundary between the first and second order domains is given by

\begin{equation}
T_{\rm up}(f) = { {2 E_1 - E_0} \over {k_B (\ln \nu - \ln f)} }, 
\label{eq:gvul12}
\end{equation}

\noindent
with the first order domain 
for temperatures
below this curve and the second order domain 
above it.
For parameters on this boundary
line the recombination efficiency is $\eta=1/2$.

Note that in  
Eq.~(\ref{eq:N})
the coverage $\rho$ is not limited to the range
$0 \le \rho \le 1$.
In the LH kinetics, the coverage is
bounded from above due to the rejection of atoms deposited
on top of other H atoms that are already adsorbed.
To include this effect, the flux term $f$ in
Eq.~(\ref{eq:N})
is replaced by
$f(1 - \rho)$. 
Solving the new equation at steady state we obtain

\begin{equation}
\eta = {1 \over 8} 
{(W+f)/a \over f/(W+f)}
\left[ -1 + \sqrt{1 + 8 {f/(W+f) \over (W+f)/a} }   \right]^2,
\label{eq:steadyetaLH}
\end{equation}

\noindent
which is similar to
Eq.~(\ref{eq:steadyeta})
except that now $W$ is replaced by $(W+f)$.
The LH rejection introduces further
constraints on the domain of high efficiency of H$_2$
formation.
This is due to the fact that at low temperature the coverage
approaches unity and newly deposited atoms are rejected.
For simplicity we will define the high efficiency domain as the
set of points in parameter space for which
$\eta \ge 1/2$.
The boundary of this domain, namely the curve 
in the ($W/a,f/W)$ plane
on which
$\eta = 1/2$,
is given by

\begin{equation}
W/a = {f/W \over (1+f/W)^2}.
\label{eq:12boundary}
\end{equation}

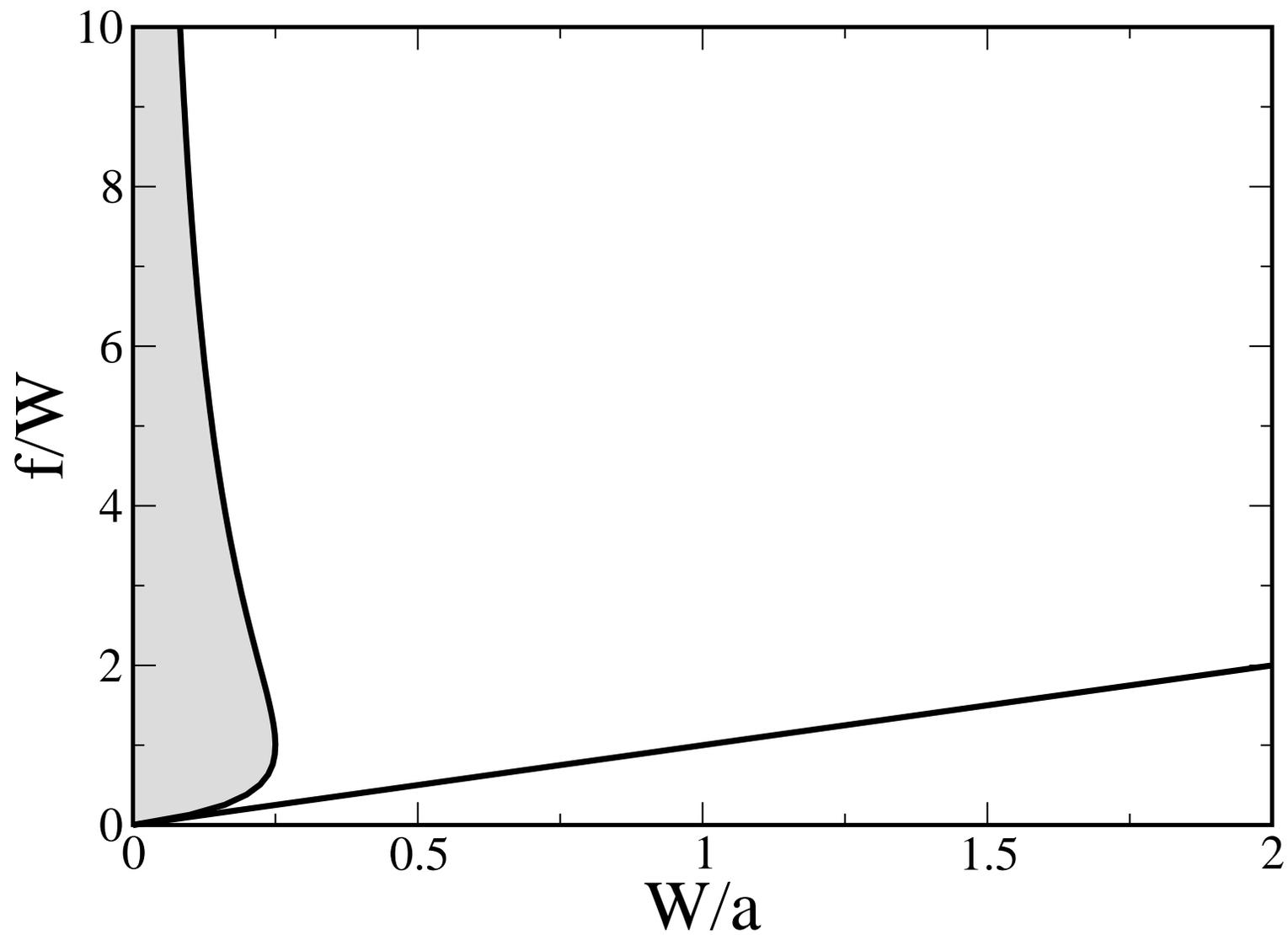
\begin{figure}
\caption{
The phase diagram of hydrogen recombination 
on macroscopic surfaces under the LH
kinetics.
The high efficiency domain is bounded to the gray area,
on the 
left side of
the curve given by
Eq.~(\ref{eq:12boundary}).
}
\label{fig:3}
\end{figure}

\noindent
In Fig. \ref{fig:3}
we show
the high efficiency 
domain (gray area)
and the low efficiency
domain 
in parameter space for the LH kinetics.
For $f/W \ll 1$
the boundary 
between them 
coincides with the diagonal
line that separates the first and second order domain without
the LH rejection.
For any given value of $W/a$ in the range
$0 < W/a < 1/4$, 
the high efficiency domain is bounded 
from above and below
by
$f_{\rm low}(W/a)/W < f/W < f_{\rm up}(W/a)/W$
where
$f_{\rm up}(W/a)$
and
$f_{\rm low}(W/a)$
are determined by 
Eq.~(\ref{eq:12boundary}).
The two boundaries are related to each other according to

\begin{equation}
{f_{\rm up}(W/a) \over W} = { W \over f_{\rm low}(W/a) }.
\end{equation}

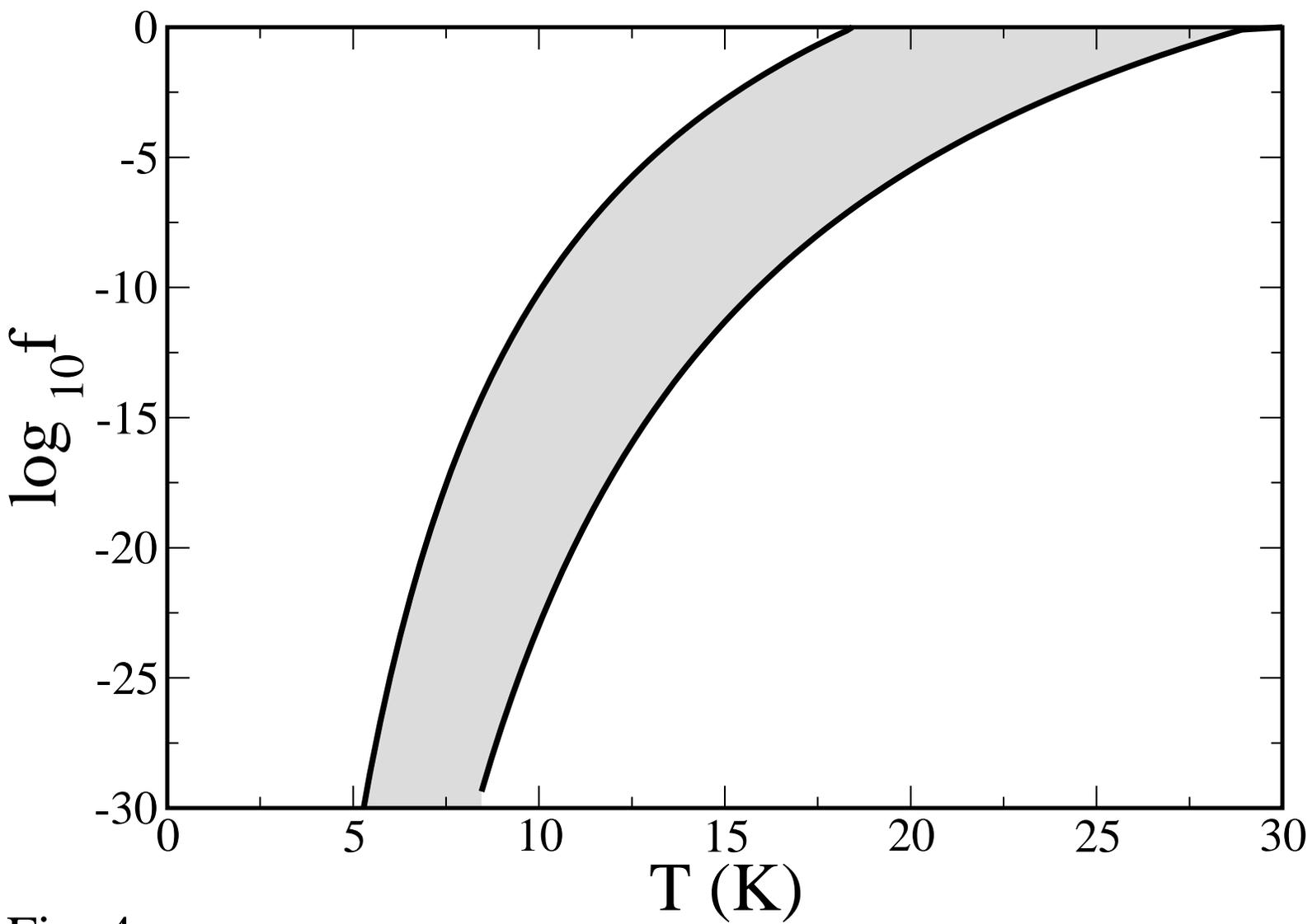
\begin{figure}
\caption{
The phase diagram 
of the hydrogen recombination process on macroscopic surfaces
in the $(T,\ln f)$
plane under the LH kinetics.
The strip of high efficiency is now bounded from both sides,
namely for any given flux there is a range of temperatures
that exhibits high recombination efficiency.  
}
\label{fig:4}
\end{figure}

For a given surface, for which $E_0$, $E_1$ and $\nu$ are known, we can
draw a diagram for the recombination process in the $(T,\log_{10}f)$ plane
(Fig. \ref{fig:4}).
For a given flux, the temperature in the domain of high efficiency
(gray area) 
is now bounded from above by
Eq.~(\ref{eq:gvul12})
and from below
by

\begin{equation}
T_{\rm low}(f) = {E_0 \over {k_B(\ln \nu - \ln f)}}.
\end{equation}

\noindent
Therefore, in order for a given surface to exhibit
a domain of high efficiency, the condition
$T_{\rm up}(f) > T_{\rm low}(f)$
must be satisfied.
This condition is satisfied if
$E_1 > E_0$, namely the activation energy for desorption is higher
than for diffusion.

A further constraint on the high efficiency domain may appear due to 
the existence of H$_2$ molecules on the surface.
These molecules may reject some of the deposited atoms through the 
LH mechanism and thus reduce the effective flux.
The H$_2$ molecules on the surface may be either molecules that
formed on the surface and  
did not desorb upon formation
\cite{Katz1999},
or ones that were adsorbed from
the gas phase.
In both cases the condition for the existence of a high efficiency
domain is that
$E_2 < 2 E_1 - E_0$,
where $E_2$ is the activation energy for desorption
of H$_2$ molecules from the surface.
Thus, surfaces that do not satisfy this condition 
(particularly surfaces on which H$_2$ molecules are adsorbed more
strongly thatn H atoms)
are
not expected to efficiently catalyze the H$_2$ formation
process,
unless molecules are desorbed
upon formation and their density in the gas phase is too low to saturate
the adsorption sites on the surface.

For small grains and low flux one may reach the situation in which
the average number of H atoms on a grain is of order unity or even less.
Under these conditions the rate equation model, 
that takes
into account only average densities, 
ignoring the fact that H atoms are discrete entities,
does not
account correctly for the recombination rate.
This is due to the fact that 
the recombination process requires at least two atoms on the
surface and
the fluctuations in the number of
H atoms on different grains become dominant. 
A more complete description
of the recombination process is needed.
Such description is provided by the master equation presented below. 

\section{H$_2$ Formation on Small Dust Grains}

\subsection{The Master Equation Model}
\label{sec:MasterModel}

We will now consider the formation of H$_2$ molecules on small
dust grains.
In this case it is more convenient to rescale the parameters
such that instead of using quantities per unit area - the total
amount per grain will be used.
The number of H atoms on the grain is 
denoted by
$n$.
Its expectation value is
given by
$\langle n \rangle = S \cdot \rho $
where $S$ is the number of adsorption sites on the grain.
The incoming flux of H atoms onto the grain surface is given by
$F  = S \cdot f $ (atoms s$^{-1}$).
The desorption rate 
$W $
remains unchanged.
The hopping rate 
$a $
(hops s$^{-1}$)
is replaced by 
$A  = a /S$
which is approximately the inverse of the time
$t_s$
required for an atom
to visit nearly all the
adsorption sites on the grain surface.
This is due to the fact that in two dimensions the 
number of distinct sites visited by a random walker
is linearly proportional to the number of steps, up
to a logarithmic correction
\cite{Montroll1965}.
The H$_2$
production rate of a single grain
is given by
$R = S \cdot r$
(molecules s$^{-1}$).
The rate equation 
will now take the form

\begin{equation}
\label{eq:Ngrain}
{ {d{ \langle n  \rangle }} \over {dt}}  =  F  
- W  \langle n  \rangle - 2 A  {\langle n  \rangle}^{2}. 
\end{equation}

\noindent
Under given flux and surface temperature, 
for grains that are large enough to hold many H atoms,
Eq.~(\ref{eq:Ngrain})
provides a good description of the recombination process.
However, for small enough grains 
$\langle n \rangle$ 
becomes or order unity
and 
Eq.~(\ref{eq:Ngrain})
becomes unsuitable,
because it neglects the fluctuations in the
number of atoms on a grain.

We will now introduce the master equation, that provides a correct
description of the recombination process even in the limit of small
grain sizes and low flux.
Consider a grain that is exposed to a flux $F$ of H atoms.
At any given time the number of H atoms adsorbed on the grain
may be $n =0, 1, 2, \dots,S$.
The probability that there are $n $ hydrogen atoms  
on the grain 
is given by
$P (n )$,
where

\begin{equation}
\sum_{n =0}^{S} P (n ) =1.
\label{eq:normalization}
\end{equation}

\noindent
The master equation provides
the time 
derivatives
of these probabilities,
$\dot P (n )$, namely the
gain or loss of the probabilities of the different
states.  These derivatives are linear in the probabilities
themselves.
The equations 
include three terms.
The first term describes the effect of the incoming flux $F$.
The probability $P (n )$ increases when an H atom is adsorbed on a grain that already
has $n -1$ adsorbed atoms 
[at a rate of $F  P (n -1)$], 
and decreases when a new atom is adsorbed on a grain with
$n $ atoms on it
[at a rate of $F  P (n )$].
The adsorption process is considered as 
completely random (Poisson process), and is fully characterized
by $F$.
The second term includes the effect of desorption. 
An H atom desorbed from a grain with $n $ adsorbed atoms, decreases the
probability $P (n )$
[at a rate of
$n  W  P (n )$, where the factor $n $ is due to the fact that each of the
$n $ atoms can desorb], 
and increases the probability
$P (n -1)$
at the same rate.
The third term describes the effect of recombination.
The production of one molecule reduces the number 
of adsorbed atoms
from $n $ to $n -2$.
For a given pair of H atoms,
the recombination rate is 
proportional to 
the sweeping rate
$A $ 
multiplied by 2 since both atoms are mobile
simultaneously.
This rate is multiplied by
the number of possible pairs of atoms, namely
$n (n -1)/2$. 
The master equation exhibits the Markov property, namely, no memory effects
are included
\cite{vanKampen1981}.
This property emerges from the fact that the incoming flux keeps washing out
any spatial correlations that may develop due to recombination events
between adjacent atoms. 
The master equation thus takes the form:

\begin{eqnarray}
\dot P (0) &=& - F  P (0) + W  P (1) 
+ 2 \cdot 1 \cdot A  P (2) \nonumber \\
\dot P (1) &=& F  \left[ P (0) - P (1) \right] 
+ W  \left[ 2 P (2) - P (1) \right] 
                + 3 \cdot 2 \cdot A  P (3) \nonumber \\
\dot P (2) &=& F  \left[ P (1) - P (2) \right] 
+ W  \left[ 3 P (3) - 2 P (2) \right] \nonumber \\
                &+& A  \left[ 4 \cdot 3 \cdot P (4) 
                 - 2 \cdot 1 \cdot P (2) \right] \nonumber \\
&\vdots& \nonumber \\
\dot P (n ) &=& F  \left[ P (n -1) - P (n ) \right] 
+ W  \left[ (n +1) P (n +1) - n  P (n ) \right] \nonumber \\
                &+& A  \left[ (n +2)(n +1) P (n +2) 
-  n (n -1) P (n ) \right]   \nonumber \\
&\vdots& \nonumber \\
\dot P (S ) &=& F  \left[ P (S -1) - P (S ) \right] 
- S W P (S ) -  S (S -1) A P (S ).    
\label{eq:Nmicro}
\end{eqnarray}

\noindent
Note that the equations for 
$\dot P (0)$ 
and
$\dot P (1)$ 
do not include all the terms, because at least one H 
atom is required for desorption to occur and at least two
for recombination.
Similarly,
the equation for 
$\dot P (S)$
does not include all the terms since there is no room 
for more than $S$ atoms 
on the grain surface.
The expectation value for the number of H atoms on the grain 
is  

\begin{equation}
\langle n  \rangle = \sum_{n =0}^{S} n  P (n ).
\label{eq:n_mean}
\end{equation}

\noindent
The rate of formation of H$_2$ molecules,
$R$ (molecules s$^{-1}$), 
is thus
given by 

\begin{equation} 
R = A  \sum_{n =2}^{S} n (n -1) P (n ).
\label{eq:Rgrain}
\end{equation}

\noindent
The hydrogen recombination efficiency on the grains
is given by

\begin{equation}
\eta = {R \over (F /2)}.
\end{equation}

\subsection{Exact Solution of the Master Equation}

When a grain is maintained
at a constant temperature 
(namely 
$W $ 
and
$A $ 
are fixed), 
and is exposed to a constant flux
$F $, 
the recombination process on its surface approaches a steady state.
Under steady state conditions the time derivatives on the left hand side
of 
Eq.~(\ref{eq:Nmicro})
is zero.
We thus obtain a homogeneous set of coupled linear equations in
the variables $P (n)$, $n=0,1,2,\dots,S$.
This set can be expressed in a matrix form as

\begin{equation}
M \vec{P} =\vec{0},
\end{equation}

\noindent
where

\begin{equation}
M =
\left( \begin{array}{cccccc}
         -F & W & 2 A & 0 & \dots & 0   \\
          F & -F - W & 2 W & 3 \cdot 2 A & \dots & 0     \\
          0 & F & - F - 2 (W + A) & 3 W & \dots & S \cdot (S-1) A     \\
          0 & 0 & F & - F - 3 (W + 2 A) & \dots & S W      \\
          \vdots & \vdots & \vdots  & \vdots & & \vdots     \\
          0 & 0 & 0 & F & \dots & - F - S [W + (S-1) A] 
         \end{array} \right)
\end{equation}

\noindent
and
$\vec P$
consists of the components
$P(n)$, $n=0,1,\dots,S$.
%
%
%
The matrix
elements are denoted
by
$M(n,m)$, $n,m=0,1,2,\dots,S$.
The only non-vanishing 
matrix elements are

\begin{eqnarray}
M(n,n)\ \ \ \ \ &=& -[F+nW+n(n-1)A] \nonumber \\
M(n+1,n) &=& F \nonumber \\
M(n,n+1) &=& (n+1)W \nonumber \\
M(n,n+2) &=& (n+2)(n+1)A. 
\end{eqnarray}

\noindent
For a finite grain the set of equations
(\ref{eq:Nmicro})
is truncated by
$n \le S$.
However, under interstellar conditions we expect low coverage
of H atoms on the grain,
namely
$\langle n \rangle \ll S$.
Therefore, one can impose a cutoff at some value $s < S$
such that 
$P(n)=0$ for $n > s$
and the normalization of probabilities now takes the form
$\sum_{n=0}^{s} P(n) =1$.
The terms in the matrix that represent flow of probabilities
between 
$P(n)$, $n \le s$
and 
$P(n)$, $n > s$
are removed.
Three of these terms disappear since they are outside the 
$(s+1) \times (s+1)$ size
matrix
that we now consider.
The term $(-F)$ in
the matrix element 
$M(s,s)$, that represents the
addition of an H atom to a grain that already includes
$s$ atoms is also removed.
The modified term will be
$M(s,s)=-s[ W + (s-1)A]$.

We proceed by performing linear operations on the rows of 
the matrix $M$.
Starting from the top, we add each element 
$M(0,m)$, $m=0,\dots,s$
of the first row
to the corresponding element 
$M(1,m)$
in the second row.
We then proceed downwards, adding
the elements of the $n$th row
to the corresponding elements of the ($n+1$)th row.
%
%
%
%
Each row of the resulting matrix includes one diagonal term
and two off-diagonal terms of the form:
$M^{\prime} (n,n)  = -F  $, 
$M^{\prime} (n,n+1)    = (n+1) (W + nA) $
and
$M^{\prime} (n,n+2)    = (n+2)(n+1)A $
(except for the last row in which all the elements are zero and the
next to the last row in which the second off-diagonal element is removed).
In order to remove the second off-diagonal terms
$M^{\prime}(n,n+2)$, 
$n=0,\dots,s-2$
we
now perform a second set of operations, this time
starting at the bottom rows.
We first 
subtract from
each element 
$M^{\prime}(s-2,m)$, 
$m=0,\dots,s$ 
of the $(s-2)$th
row, the corresponding element 
$M^{\prime}(s-1,m)$, 
of the $(s-1)$th
row, multiplied by 
$M^{\prime}(s-2,s)/M^{\prime}(s-1,s)$.
We then proceed in a similar fashion all the way up.
Each line in the resulting matrix has only
one diagonal and one off-diagonal term.
The diagonal elements take the form
$M^{\prime\prime}(n,n) = -F$,
$n = 0,\dots,s-1$
and 
$M^{\prime\prime}(s,s) = 0$.
The off-diagonal elements will be

\begin{eqnarray}
M^{\prime\prime} (s-1,s)\ \ \ \ \ \ \ \   \ \ \  &=& M^{\prime} (s-1,s) \nonumber \\ 
M^{\prime\prime} (s-2,s-1) \ \ \ \ \ \  &=& M^{\prime} (s-2,s-1) 
- { M^{\prime} (s-2,s) 
   \over
   M^{\prime\prime} (s-1,s) } 
\cdot    M^{\prime} (s-1,s-1)  \nonumber \\
M^{\prime\prime} (s-3,s-2) \ \ \ \  \  &=& M^{\prime} (s-3,s-2)  
- { M^{\prime} (s-3,s-1) 
   \over
   M^{\prime\prime} (s-2,s-1) } 
\cdot    M^{\prime} (s-2,s-2)  \nonumber \\
\vdots \nonumber \\
M^{\prime\prime} (s-n,s-n+1) &=& M^{\prime} (s-n,s-n+1) \nonumber \\
&-& { M^{\prime} (s-n,s-n+2) 
   \over
   M^{\prime\prime} (s-n+1,s-n+2) } 
\cdot    M^{\prime} (s-n+1,s-n+1)  \nonumber \\
\vdots \nonumber \\
M^{\prime\prime} (0,1) \ \ \ \ \ \ \ \ \ \ \ \ \ \ \ \  &=& M^{\prime} (0,1) 
- { M^{\prime} (0,2) 
   \over
   M^{\prime\prime} (1,2) }
\cdot M^{\prime\prime} (1,1)  
\label{eq:pruning}
\end{eqnarray}

\noindent
By combining the operations in
Eq.~(\ref{eq:pruning})
we express the non-vanishing off-diagonal elements 
of the matrix
$M^{\prime\prime}$
as continued fractions.
Their reduced form is

\begin{equation}
{ M^{\prime\prime}(n,n+1) \over {(n+1)\sqrt{AF}} }  = 
\sqrt{A \over F} ({W \over A}+n) +{1 \over {\sqrt{A \over F} ({W \over A}+n+1) + {1 \over {
\cdot \cdot \cdot
+ {1 \over {\sqrt{A \over F} ({W \over A}+s)}}}} }}
\label{eq:contfrac}
\end{equation}

\noindent
for $n=0,\dots,s-1$.
The equation
$M^{\prime\prime} \vec P = \vec 0 $
now takes the form of
a set of recursion equations: 

\begin{equation}
P(n+1) = {- M^{\prime\prime}(n,n)\over M^{\prime\prime}(n,n+1)} P(n), \ \ \ n=0,\dots,s-1. 
\end{equation}

\noindent
Using these equations we can express all the probabilities
in terms of $P(0)$ according to

\begin{equation}
P(n) = (-1)^n P(0) \prod_{i=0}^{n-1} \left[{ M^{\prime\prime}(i,i) \over  M^{\prime\prime}(i,i+1) }\right] , 
\ \ \ n=1,\dots,s. 
\label{eq:exactP}
\end{equation}

\noindent
The probability $P(0)$ is then determined by
the normalization condition
$\sum_{n=0}^{s} P(n) =1$.
%
%
%
%
Eq.
(\ref{eq:exactP}),
complemented by the normalization condition,
provides an exact solution
of the master equation under steady state conditions
for any finite cutoff $s \le S$.
Since the master equation is of use when the coverage is very
low, the tail of $P(n)$ already decays for some $n \ll S$.
Therefore, in evaluating $P(0)$ it is sensible to ignore the
cutoff at $s \le S$ and write the infinite sum instead.
In this case, 
the solution of the master equation can be
expressed in terms of Bessel functions.
The connection to the Bessel functions can be obtained
from the (infinite) continued fraction expression in 
Eq.~(\ref{eq:contfrac}), obtained when $s \rightarrow \infty$.
Using the continued fraction expansion of the
ratio 
$J_{\nu}(z)/J_{\nu-1}(z)$
in Ref. 
\cite{Abramowitz}
(Eq. 9.1.73 in page 363)
and the relation
$I_{\nu}(z)/I_{\nu-1}(z) = - i J_{\nu}(iz)/J_{\nu-1}(iz)$
we obtain that

\begin{equation}
{ M^{\prime \prime}(n,n+1)
\over
{(n+1) \sqrt{FA}} }
=
{ {I_{W/A+n-1}\left(2 \sqrt{F/A}\right)}
\over
{I_{W/A+n}\left(2 \sqrt{F/A}\right)}.
}
\end{equation}

\noindent
We thus obtain an expression for
$P(n)$ of the form

\begin{equation}
P(n) = {1 \over n!} P(0) \left(\sqrt{F \over A}\right)^{n} 
{ I_{W/A+n-1} \left(2\sqrt{F/A}\right) \over I_{W/A-1} \left(2\sqrt{F/A}\right)}. 
\label{eq:exactPBessel}
\end{equation}

%
%
%
%

\noindent
The normalization factor $P(0)$
can be expressed in terms of Bessel functions,
using 
Ref. 
\cite{Abramowitz}
(Eq. 9.6.51 in page 377, with $\lambda=\sqrt{2}$), as

\begin{equation}
P(0) = { 2^{{1 \over 2} ({W \over A} -1)}{I_{W/A-1} \left(2\sqrt{F/A}\right)} 
\over 
{I_{W/A-1} \left(2\sqrt{2 F/A}\right)} }.  
\label{eq:P0B}
\end{equation}

\noindent
Therefore,

\begin{equation}
P(n) = { 2^{{1 \over 2} ({W \over A} -1)} \over n!}
\left( \sqrt{F \over A}\right)^{n}
{ {I_{W/A+n-1} \left(2\sqrt{F/A}\right)} 
\over 
{I_{W/A-1} \left(2\sqrt{2 F/A}\right)}} ,  
\label{eq:PnB}
\end{equation}

\noindent
in agreement with
Ref.
\cite{Green2001},
where the solution was obtained using a generating function.
To examine the effect of the cutoff at $s \le S$, we compared
the distributions $P(n)$ obtained from 
Eqs.~(\ref{eq:exactP}) 
for different values of  
$s$, as well as the distribution expressed in terms of the Bessel
functions for which
$s \rightarrow \infty$.
Under the conditions of low coverage studied here
we observe a very fast convergence of
Eq.~(\ref{eq:exactP}) 
to 
Eq.~(\ref{eq:PnB}) 
as $s$ increases.
We have also performed direct numerical integration of the
master equation and found that the solution
given by
Eq.~(\ref{eq:exactP}) 
is stable and the convergence of the integration process 
is fast.

Using similar summations,
we can now find exact expressions for the first and
second moments of the distribution $P(n)$. 
The average number of H atoms on the grain
is given by

\begin{equation}
\langle n \rangle = \sqrt{F \over {2A}} 
{ {I_{W/A} \left(2\sqrt{2 F/A}\right)} 
\over 
{I_{W/A-1} \left(2\sqrt{2 F/A}\right)} }.  
\label{eq:n_avB}
\end{equation}

\noindent
The rate of formation of H$_2$ molecules is given by
$R = A \langle n(n-1) \rangle$, where

\begin{equation}
\langle n(n-1) \rangle = 
{F \over {2A}} 
{ {I_{W/A+1} \left(2\sqrt{2 F/A}\right)} 
\over 
{I_{W/A-1} \left(2\sqrt{2 F/A}\right)} },  
\label{eq:nnm1_avB}
\end{equation}

\noindent
and the recombination efficiency
is

\begin{equation}
\eta = 
{ {I_{W/A+1} \left(2\sqrt{2 F/A}\right)} 
\over 
{I_{W/A-1} \left(2\sqrt{2 F/A}\right)} }.  
\label{eq:eta_avB}
\end{equation}

\noindent
The second moment 
$\langle n2 \rangle$ 
can be obtained
as the sum of the right hand sides of 
Eqs.~(\ref{eq:n_avB})
and
(\ref{eq:nnm1_avB}).
The fluctuations in the number of H atoms on different grains 
can be quantified by the standard deviation of the 
distribution $P(n)$, given by

\begin{equation}
\sigma = \sqrt{\langle n^2 \rangle - 
\langle n \rangle^2}.
\label{eq:sigma} 
\end{equation}

The limits of first and second order kinetics  
for large grains can now be
reproduced from 
Eq.~(\ref{eq:eta_avB}).
The extreme limit of first order kinetics 
is characterized by negligible desorption, 
namely 
$W/A \ll 1$
and for large enough grains this limit
also satisfies
$F/W \gg 1$.
In this limit the indices of the Bessel functions
in 
Eq.~(\ref{eq:eta_avB})
$W/A \pm 1 \rightarrow \pm 1$,
respectively,
and
due to the symmetry
$I_{-n}(z) = I_n(z)$
of the Bessel functions
\cite{Abramowitz}
(Eq. 9.6.6 in page 375)
$\eta \rightarrow 1$,
in agreement with the rate equations.
The extreme limit of second order kinetics on large 
grains is characterized by
$F/W \ll W/A$.
Assuming that the geometric mean of $F/W$ and $W/A$
satisfies
$\sqrt{F/A} \ll 1$ we obtain

\begin{equation}
\eta = {2 F \over W} {1 \over {\left( {W \over A}-1\right)}}.
\label{eq:etaapprox}
\end{equation}

\noindent
In the case of very large grains, namely $W/A \gg 1$
we thus obtain $\eta = 2AF/W2 = 2af/W2$ which
is in agreement with the rate equation results. 

Using 
Eqs.
(\ref{eq:n_mean})
and
(\ref{eq:Nmicro})
to express the time derivative of
$\langle n  \rangle$
we obtain

\begin{equation}
{d \langle n  \rangle \over dt} =
f - W \langle n  \rangle - 2 A \langle n(n-1) \rangle.
\label{eq:Nratemas}
\end{equation}

\noindent
This equation resembles the rate equation
(\ref{eq:Ngrain})
except for the recombination term in which
$\langle n  \rangle2$
was replaced by
$\langle n2 \rangle - \langle n  \rangle$.
In the limit of
small grains, 
where 
$\langle n  \rangle$
is small 
while the fluctuations
represented by 
$\sigma$ become dominant,
the rate equation becomes unsuitable and over-estimates
the rate of H$_2$ production.

\subsection{Analysis and Results}

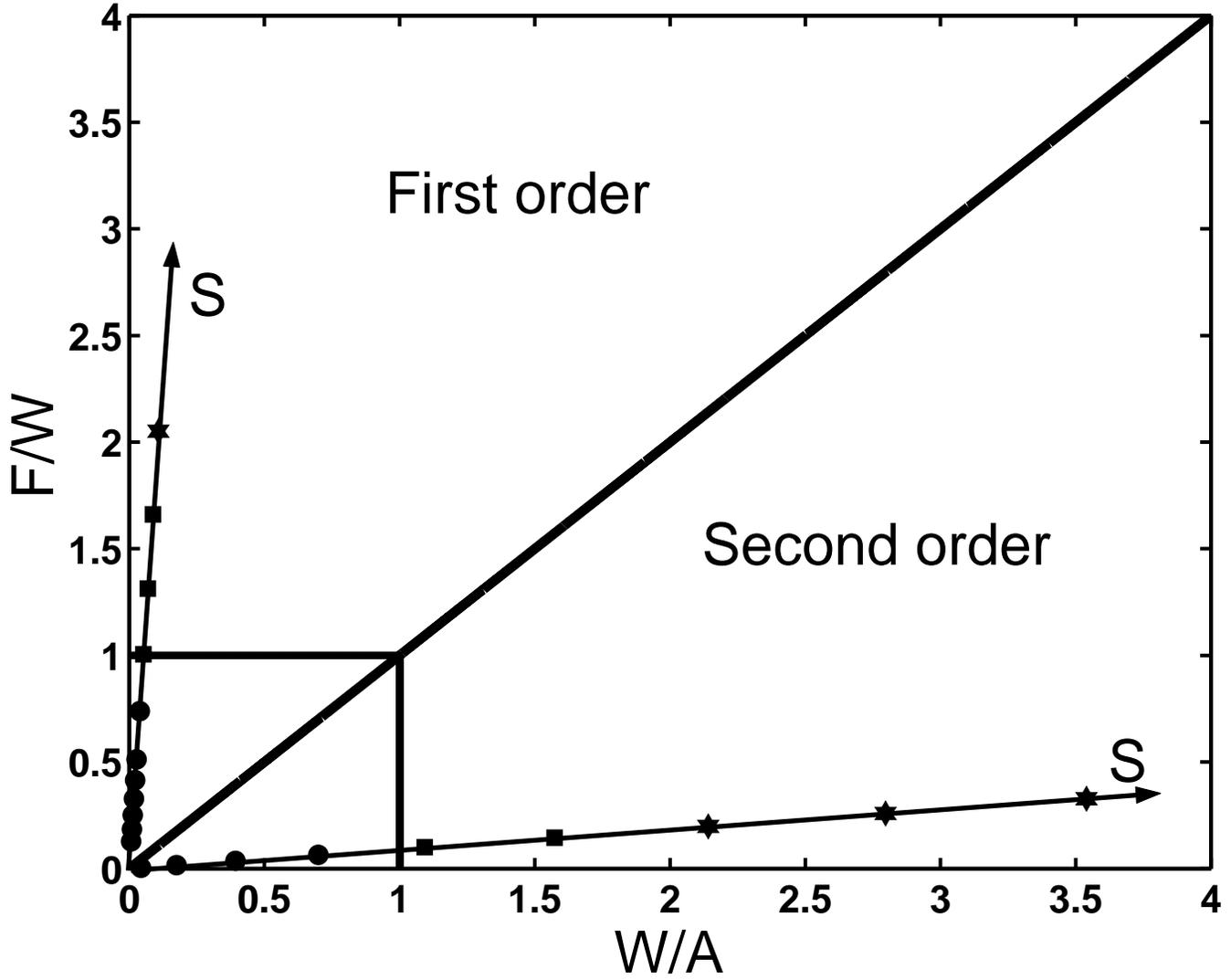
\begin{figure}
\caption{
The phase diagram of the hydrogen recombination process on
grains 
without LH rejection,
in terms of the dimensionless parameters
$W/A$ and $F/W$. 
The diagonal
line separates between the domain of first order 
(on the left) and second order recombination (on the right),
respectively.
The (unit) square near the origin is the domain in which the
hydrogen recombination efficiency on grains (obtained from the
master equation) deviates significantly from the rate equation 
results.
In the case of first order processes, the condition
for such deviation is $F/W < 1$, 
while in second order processes
the condition is
$W/A < 1$.
Also included are two axes, in the first and
the second order domains, 
that under given physical conditions 
(fixed values of $f$ and $T$) 
represent the variation in the grain size.
The behavior of $\eta$, $\langle n \rangle$ and $\sigma/\langle n \rangle$
along these axes
is shown in Figs.
\ref{fig:6}
and
\ref{fig:7}
for the first and second order cases, respectively.
}
\label{fig:5}
\end{figure}

Consider the recombination process on a small grain with $S$ adsorption sites
under steady state conditions.
The boundary between the first order and the second order regimes 
can be expressed in terms of
$F = f \cdot S$
and
$A = a/S$,
taking the form
$F/W = W/A$
(neglecting the LH rejection).
In the first order regime
$F/W \gg W/A$,
while in the second order regime
$F/W \ll W/A$.
The domains of first and second order kinetics are shown in Fig. \ref{fig:5},
separated by the diagonal line
$F/W = W/A$.

The finite size of the grain introduces a third length scale to the problem.
We will now examine how small the grain should be in order for the 
recombination efficiency to deviate significantly from the rate equation
results.
To this end we calculate $\eta$
as a function of the grain size $S$,
using the exact solution of the master equation, presented above.
The efficiency $\eta$ 
vs. $S$
in the case of first order kinetics is shown in Fig. \ref{fig:6}(a).
The recombination efficiency obtained from the master equation (solid line)
coincides with the rate equation result (dashed line) for large grains but
declines below some grain size.
In the first order case, such deviations typically occur only for extremely
small grains of a few thousands adsorption sites. 
The average number of atoms on the grain vs. S is shown in Fig. \ref{fig:6}(b).
The $S$ axis in Figs. \ref{fig:6} corresponds to the arrow drawn in the first order
domain of Fig. \ref{fig:5}.
Identifying the corresponding symbols, we observe that the significant
decline in $\eta$ starts when 
$F/W < 1$, namely when
$S < W/f$.
The fluctuations in the number of H atoms on a grain 
can be quantified by the standard deviation 
$\sigma$ of the distribution $P(n)$, given
by Eq.~(\ref{eq:sigma}).
The standard deviation $\sigma$,
divided by
$\langle n \rangle$,
is shown 
in Fig. \ref{fig:6}(c) as a function of  
the grain size $S$.
Clearly, as the grain size decreases, the fluctuations become more pronounced.

Using the notation introduced above, in the domain of first order
kinetics the rate equation results start to deviate from the correct
value of $\eta$
for grain sizes that satisfy

\begin{equation}
S < s_{\rm vacant} < s_{\rm visit}.
\end{equation}

\noindent
Under these conditions
there is typically no atom or only a single H atom on the grain, which is 
unlikely to find a second atom to recombine with.
It thus desorbs before it has a chance to recombine.

The efficiency $\eta$ in the case of second order kinetics is shown in 
Fig. \ref{fig:7}(a).
Again, the master equation efficiency (solid line) coincides with the
rate equation result (dashed line) for large grains but
declines below some grain size,
typically of a few tens of thousands of adsorption sites. 
The average number of atoms on the grain vs. S is shown in Fig. \ref{fig:7}(b).
The deviations between the rate equations and the master equation 
results are accompanied by large fluctuations in the number
of atoms on a grain, 
as can be seen in Fig.
\ref{fig:7}(c).
The $S$ axis in Figs.  
\ref{fig:7}
corresponds to the arrow drawn in the second order
domain of Fig. 
\ref{fig:5}.
Identifying the corresponding symbols, we observe that the significant
decline in $\eta$ starts when 
$W/A < 1$, namely
$S < a/W$.
Thus, in the domain of second order kinetics, deviations between the rate
equations and the master equation occur for a range of grain sizes given by 

\begin{equation}
S < s_{\rm visit} < s_{\rm vacant}. 
\end{equation}

\noindent
In this case, the grain surface area is smaller than the area that an atom
can scan before it desorbs.
As a result the atom tends to perform several sweeps of the grain surface
visiting again the same vacant sites it has already visited.
The probability to find a second atom in these sites is much lower than
predicted by
the rate equation that does not include such return visits.
The recombination efficiency is thus sharply reduced as the grain size
further decreases.

In conclusion, we observe that in both the first and second order kinetics
the recombination efficiency on a grain (given by the
master equation) starts to deviate from the rate equation result when the
grain size (represented by $S$) becomes the smallest length scale in the 
problem.
In the first order case this happens when
$S < s_{\rm vacant}$
while in the second order case it happens when
$S < s_{\rm visit}$.

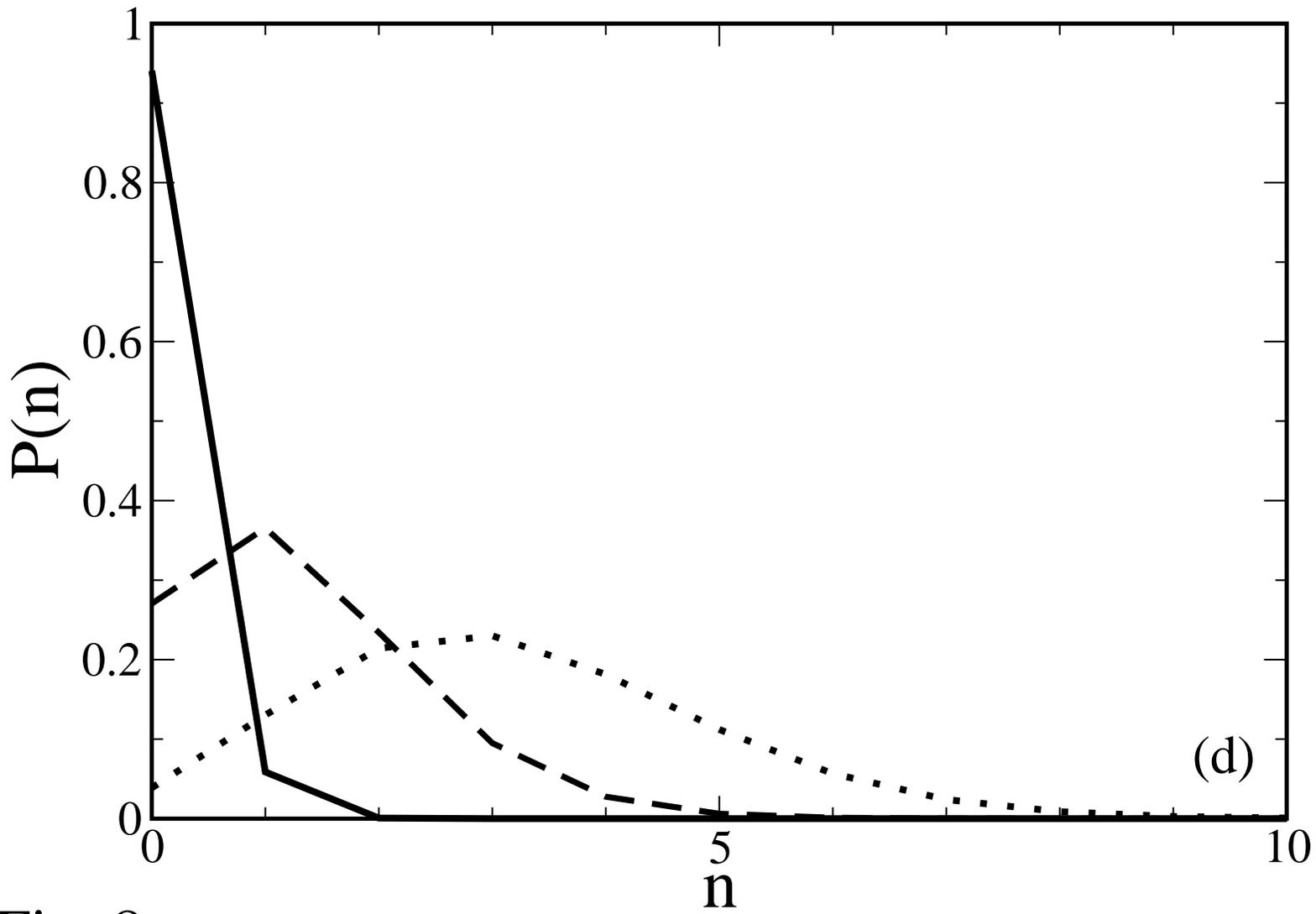
\begin{figure}
\caption{
The distribution $P(n)$ at three points along the arrow in the second order domain of
Fig. \ref{fig:5},
namely for a small grain with $S=2513$ (solid line),
and medium size grain with $S=56705$ (dashed line)
and for a large grain with
$S=141372$.
}
\label{fig:8}
\end{figure}

The distribution $P(n)$ at three points
along the $S$ axis, in the second order domain, are shown in Fig.
\ref{fig:8}.
For a relatively large grain, $P(n)$ exhibits
a well defined and nearly symmetric peak.
For a very small grain it becomes a a monotonically decreasing
function, dominated by $n=0$, 1 and 2. 

\section{Discussion}

The analysis above 
can be related to
the modified rate
equations studied in Refs.
\cite{Caselli1998,Shalabiea1998,Stantcheva2001}. 
The modification is needed when the grain size becomes smaller than
the two length scales involved in the recombination process. 
In the case of first order kinetics this occurs when
$S < W/f < a/W$,
while in second order desorption it occurs when
$S < a/W < W/f$.

Consider the case of first order kinetics. When the grain becomes smaller
than $W/f$ the typical number of atoms on the grain is smaller than one
(even if we consider the depletion of the H atoms on the surface due to 
desorption alone and neglect recombination).
Therefore, the rate in which atoms find each other on the surface is not
determined anymore by the hopping rate $a$, or the corresponding length scale $a/W$,
but by the grain size $S$. Therefore, the recombination term in 
Eq.~(\ref{eq:Ngrain})
should be modified to reflect the change 
$a/W \rightarrow S$, 
or equivalently
$A/W \rightarrow 1$. 
This is achieved by replacing $A$ by $W$.

In the case of second order kinetics, when the grain size becomes smaller
than $a/W$ the atom is typically able to perform more than a full sweep of the
entire grain before it desorbs. However, when an atom visits the same sites for
the second time, the probability of finding another H atom there is greatly 
reduced. Therefore, the recombination rate is determined by S rather than by
$a/W$, requiring the modification
$a/W \rightarrow S$, 
or the replacement of $A$ by $W$ in 
Eq.~(\ref{eq:Ngrain}),
exactly as in the case of first order kinetics.
The modified rate equations thus takes the form

\begin{equation}
\label{eq:Ngrainmod}
{ {d{ \langle n  \rangle }} \over {dt}}  =  F  
- W  \langle n  \rangle - 2 g\left({A \over W},{W \over F}\right)  
{\langle n  \rangle}^{2}. 
\end{equation}

\noindent
where 

\begin{equation}
g\left({A \over W},{W \over F}\right) = 
\left\{
\begin{array}{ll}
A:               & {A \over W} < 1 \ \ or \ \  {W \over F} <1         \\
W:               & {A \over W} >1  \ \ and \ \  {W \over F} > 1.
\end{array}
\right.
\label{limitsmod}
\end{equation}

\noindent
The modified rate equations takes into account correctly
the length scales involved in the recombination process
on small grains. It provides results for $\langle n \rangle$
that are in significantly better agreement with the master equation,
compared to the unmodified rate equations.
It also improves the results for the recombination efficiency.
However, the modified rate equations still  
involve only the average number of atoms
on a grain and 
do not take into account 
the discreteness of the H atoms and
the fluctuations, 
quantified by the second moment
in Eq.~(\ref{eq:Nratemas}). 
Since the fluctuations
dominate the recombination process on small grains,
the results of the modified rate equations for $\eta$
are not expected to coincide
with those of the master equation,
but only to approximate them better than the ordinary rate equations.  
We observe that the deviations between the rate equation results 
and the correct results obtained from the master equation
are significant mostly in the second order domain
[Fig. \ref{fig:7}].
In the domain of first order kinetics such deviations occur only for extremely 
small grains, that may be physically irrelevant
[Fig. \ref{fig:6}].

While the results of the calculations above were presented using dimensionless
parameters, the actual calculations in Figs. 
\ref{fig:6} and \ref{fig:7} were done for physically
relevant parameters. In both of them we used the parameters of the amorphous
carbon sample, measured experimentally in 
Refs.~\cite{Pirronello1999,Katz1999}.
On this sample the activation energies for H diffusion and desorption
were found to be
$E_0=44.0$ meV
and 
$E_1=56.7$ meV,
respectively.
The density of adsorption sites on the amorphous
carbon surface was found to be
$s_{dens} \cong 5 \times 10^{13}$ (sites cm$^{-2}$).
The results for first order kinetics (Fig. \ref{fig:6})
were obtained for $T=17$ K
and $f=5 \times 10^{-8}$ ML s$^{-1}$.
The results for second order kinetics (Fig. \ref{fig:7})
were obtained for $T=18$ K
and $f=3.4 \times 10^{-9}$ ML s$^{-1}$.
The connection with the density and temperature of the hydrogen in the
gas phase is made through 
$f = \rho_{\rm gas} v_{\rm gas}/4 s_{dens}$
where 
$\rho_{\rm gas}$ (atoms cm$^{-3}$)
is the density of H atoms in the gas phase,
$v_{\rm gas}$
is the typical velocity of these atoms
and the factor or 4 in the denominator is the ratio between the surface
area and the cross section for a spherical grain
\cite{Biham2001}.
The number of adsorption sites on a spherical grain of diameter
$d$ is given by

\begin{equation}
S= 4 \pi \left({d \over 2} \right)^2 \cdot s_{\rm dens},
\end{equation}

\noindent
and the flux $F= \pi (d/2)^2 \rho_{\rm gas} v_{\rm gas}$.
Observations indicate that 
the population of carbonaceous and silicate grains in interstellar 
clouds exhibits 
a broad distribution
of grain sizes, roughly in the range
$10^{-6} {\rm cm} < d < 10^{-4} {\rm cm}$
\cite{Mathis1977,Mathis1996,O'Donnell1997,Weingartner2001}

\section{Summary}

We have studied the
process of hydrogen recombination on small grains in the interstellar medium
using the master equation approach.
An exact solution of the master equation under steady state conditions was presented.
This solution provides the probability distribution
$P(n)$ of having $n$ atoms on the grain
as a function of the grain size, flux, temperature and the
parameters of the surface.
From this distribution one can obtain an exact expression for the
hydrogen recombination rate on the grain surface.
The results were compared to those obtained from the rate equations
that describe the recombination process on
macroscopic surfaces. 
In the case of a macroscopically large surface, two length scales are
identified.
One length scale is related to the average number of vacant sites
around each adsorbed atom, given by 
$s_{\rm vacant}$.
The other length scale is given by the average number of sites, 
$s_{\rm visit}$,
that an adsorbed 
atom visits before it desobs (neglecting recombination).
The relation between these two (length) scales determines the properties
of the recombination process, dividing the parameter space into two domains.
The domain 
$s_{\rm visit} > s_{\rm vacant}$ 
is characterized by 
first order kinetics with
high recombination efficiency,
while the domain
$s_{\rm visit} < s_{\rm vacant}$ 
exhibits second order kinetics 
with low recombination efficiency.
In both domains the finite size of the
grain enters as a significant factor 
(requiring the use of the master equation rather than the rate equations)
when it becomes the smallest (length) scale in the
system.
In the domain of first order kinetics the grain size becomes the
smallest (length) scale when
$S < s_{\rm vacant}$ 
while in the second order it occurs when
$S < s_{\rm visit}$. 

Acknowledgements: 
We thank I. Furman, E. Herbst, H. Mehl, V. Pirronello,
A. Schiller and G. Vidali
for helpful discussions.
This work was supported by the 
Adler Foundation for Space Research of the 
Israel Science Foundation.

\newpage

\begin{figure}
\caption{
(a) The coverage $\rho$ 
given by 
Eq.~(\ref{eq:steadyrho})
as a function of
the angle $\theta$, given by 
Eq.~(\ref{eq:theta})
(see Fig. \ref{fig:1}),
without LH rejection (solid line)
and with LH rejection (dashed line).
The results shown are for points along the curve
given by
Eq. (\ref{eq:hyperbole}). 
(b) The recombination efficiency 
$\eta$ [Eq.~(\ref{eq:steadyeta})]
vs. $\theta$, 
without LH rejection (solid line)
and with LH rejection (dashed line),
along the same curve.
Note that without LH rejection $\eta$
depends only on $\theta$, namely, it is independent
of the specific curve.
}
\label{fig:2}
\end{figure}

\begin{figure}
\caption{
The hydrogen recombination efficiency $\eta$ (a)
and the average number of atoms on the grain $\langle n \rangle$ (b)
for first order kinetics
as a 
function of the number of adsorption sites, $S$,
on the grain.
The solid line (with the symbols on it) shows the results obtained
from the master equation, and the dashed line shows the rate equation
results.
The symbols correspond to those in 
Fig. \ref{fig:5}, 
and indicate that the
deviations between the master equation and the rate equations
become significant below
$F/W=1$.
(c) The standard deviation 
$\sigma$
of the distribution $P(n)$ (normalized by 
$\langle n \rangle$). It increases sharply as the grain size is reduced
entering the range in which
$F/W<1$, indicating that in this range fluctuations play an important
role.
}
\label{fig:6}
\end{figure}

\begin{figure}
\caption{
The hydrogen recombination efficiency $\eta$ (a)
and the average number of atoms on the grain $\langle n \rangle$ (b)
for second order kinetics
as a 
function of the number of adsorption sites, $S$, 
on the grain.
The solid line (with the symbols on it) show the results obtained
from the master equation, and the dashed line show the rate equation
results.
The symbols correspond to those in 
Fig. \ref{fig:5}, 
and indicate that the
deviation between the master equation and the rate equations
become significant below
$W/A=1$.
(c) The standard deviation $\sigma$
of the distribution $P(n)$ (normalized by 
$\langle n \rangle$). It increases sharply as the grain size is reduced
entering the range in which
$W/A<1$, indicating that in this range fluctuations play an important
role.
}
\label{fig:7}
\end{figure}

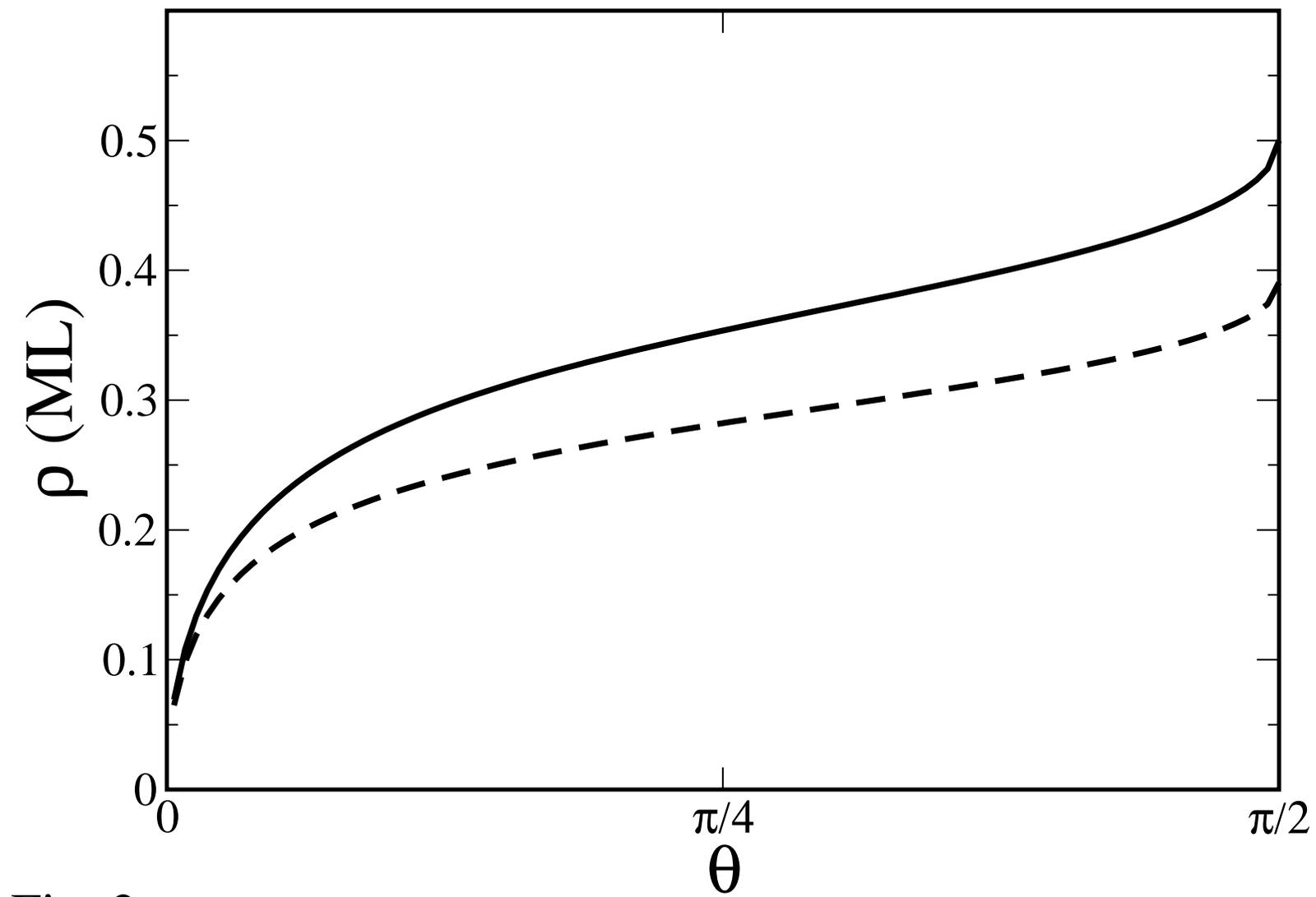
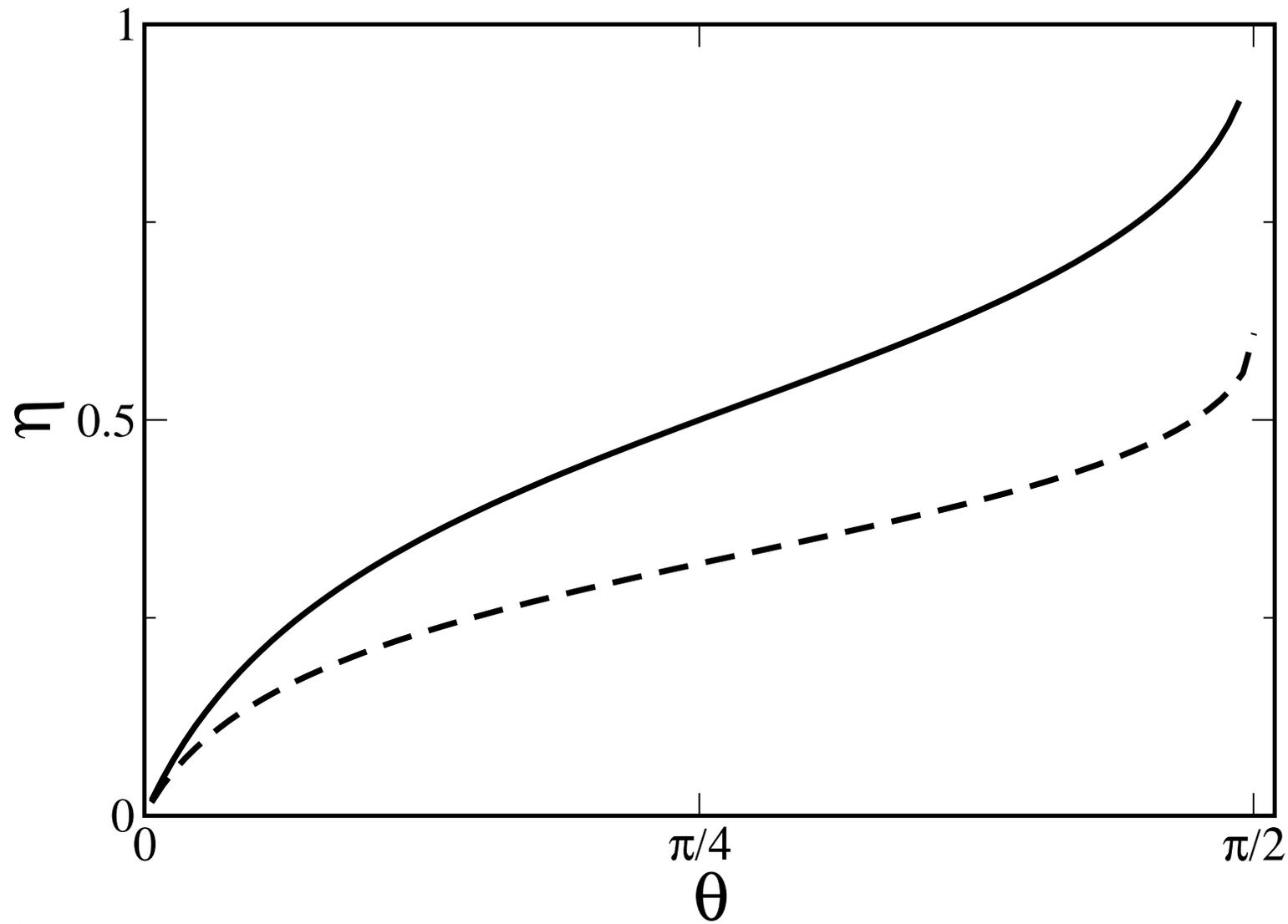
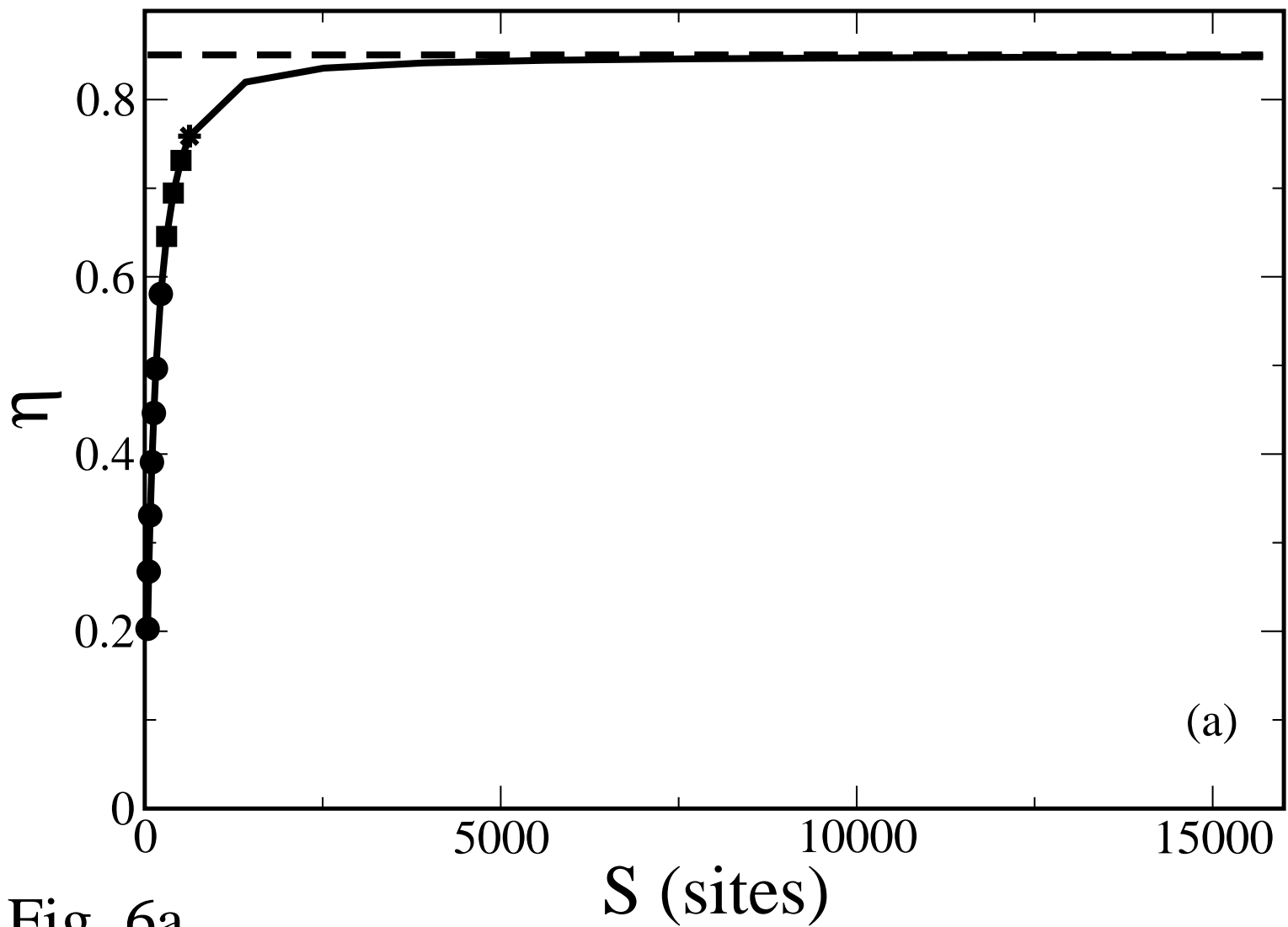
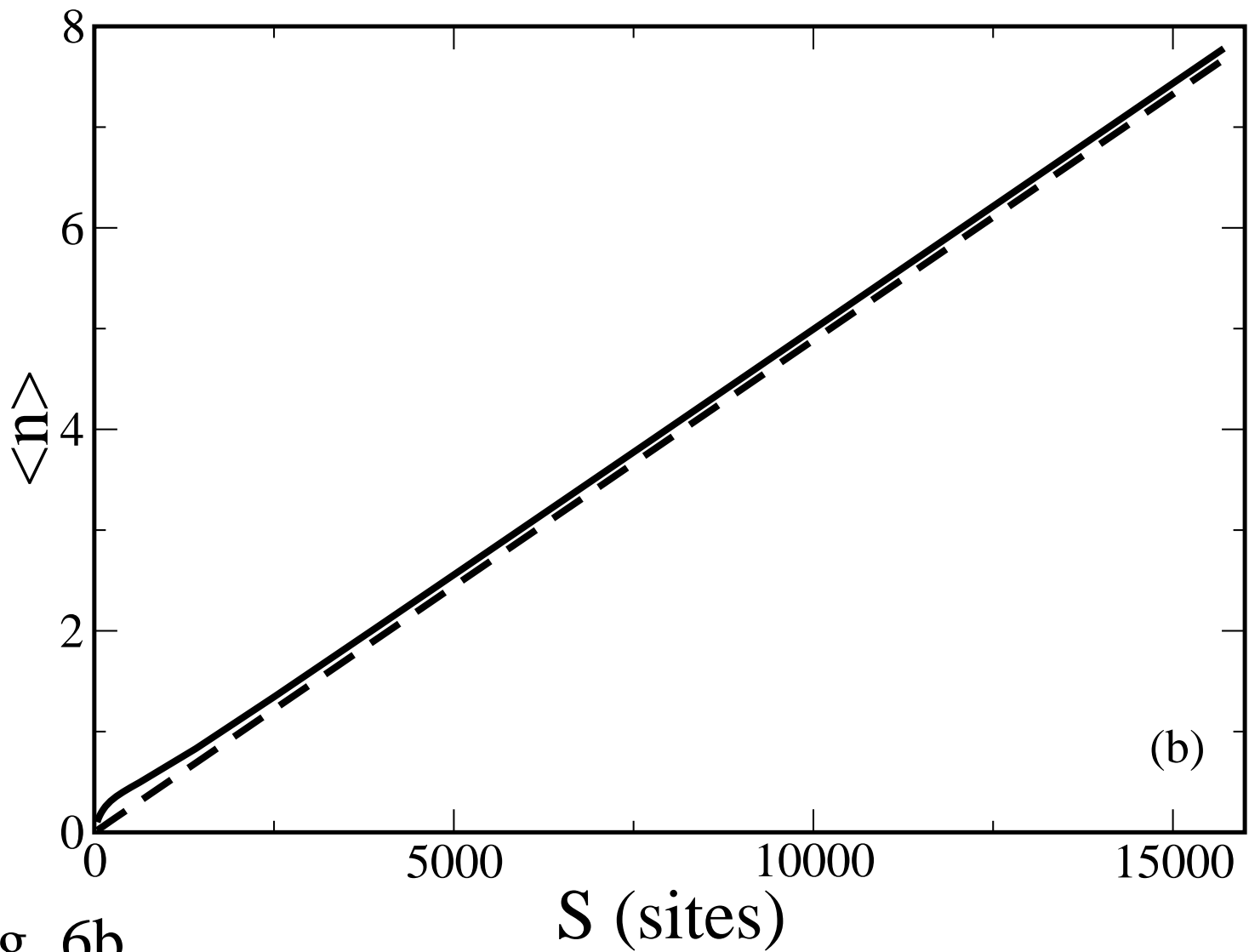
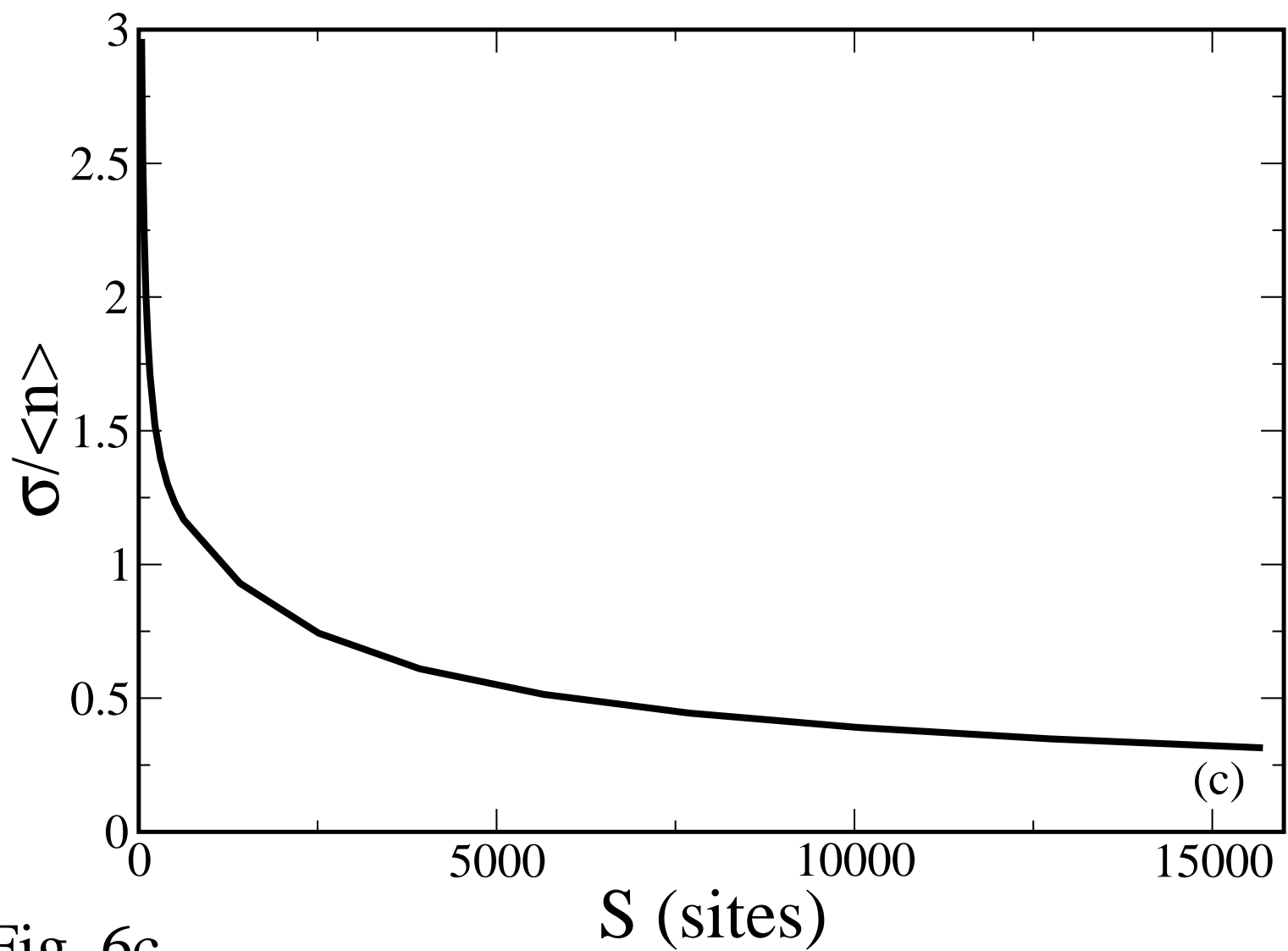
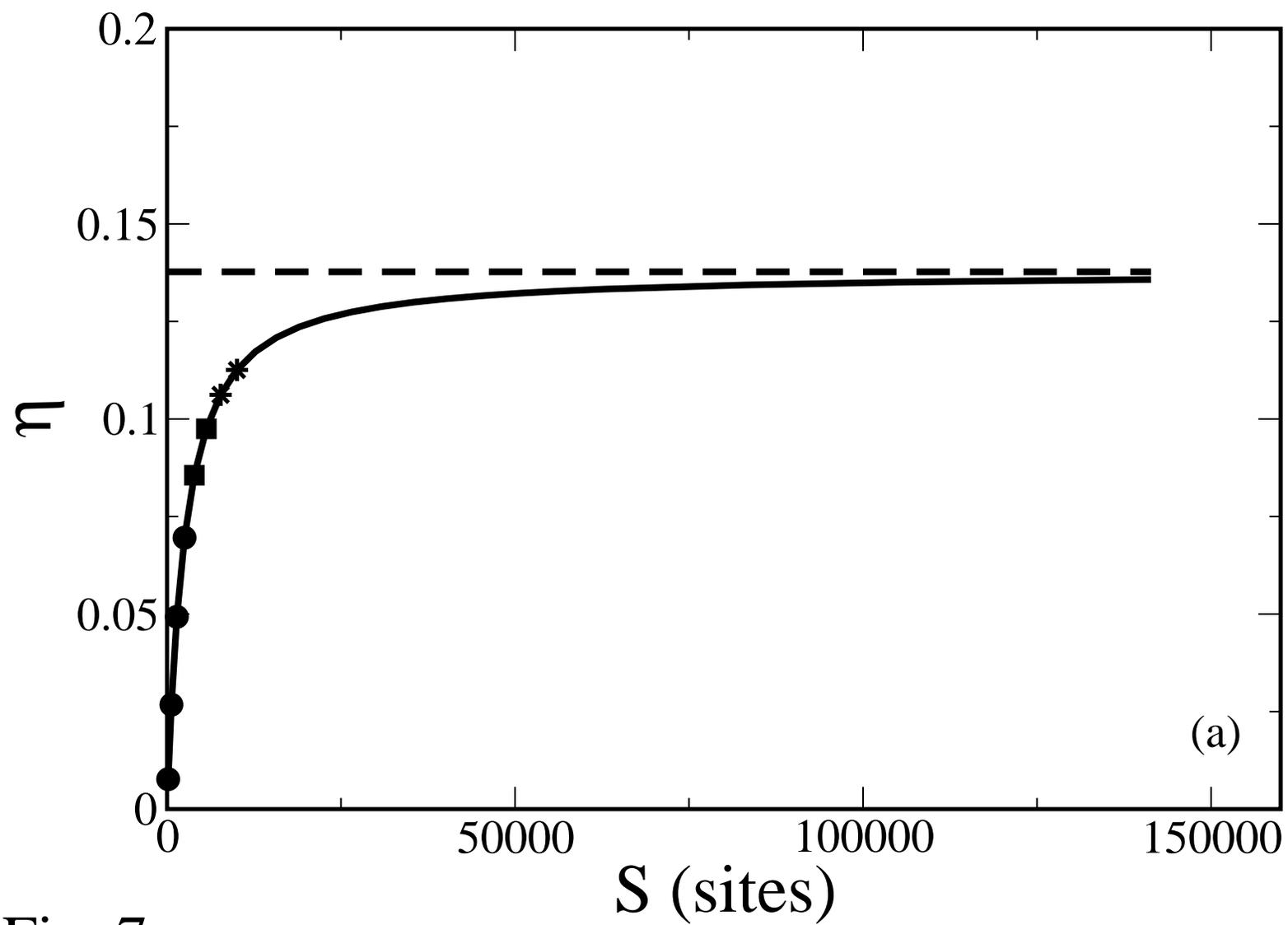
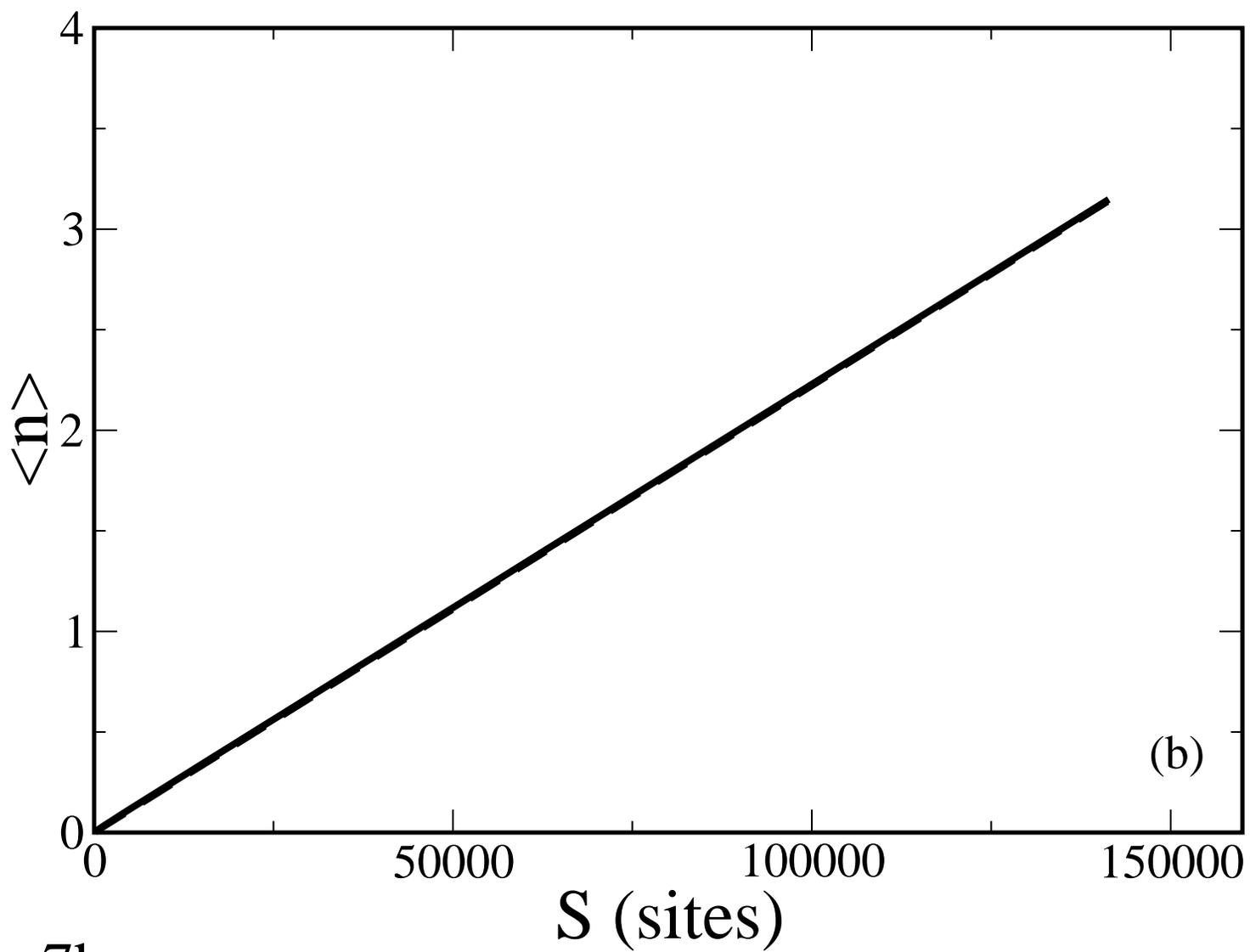
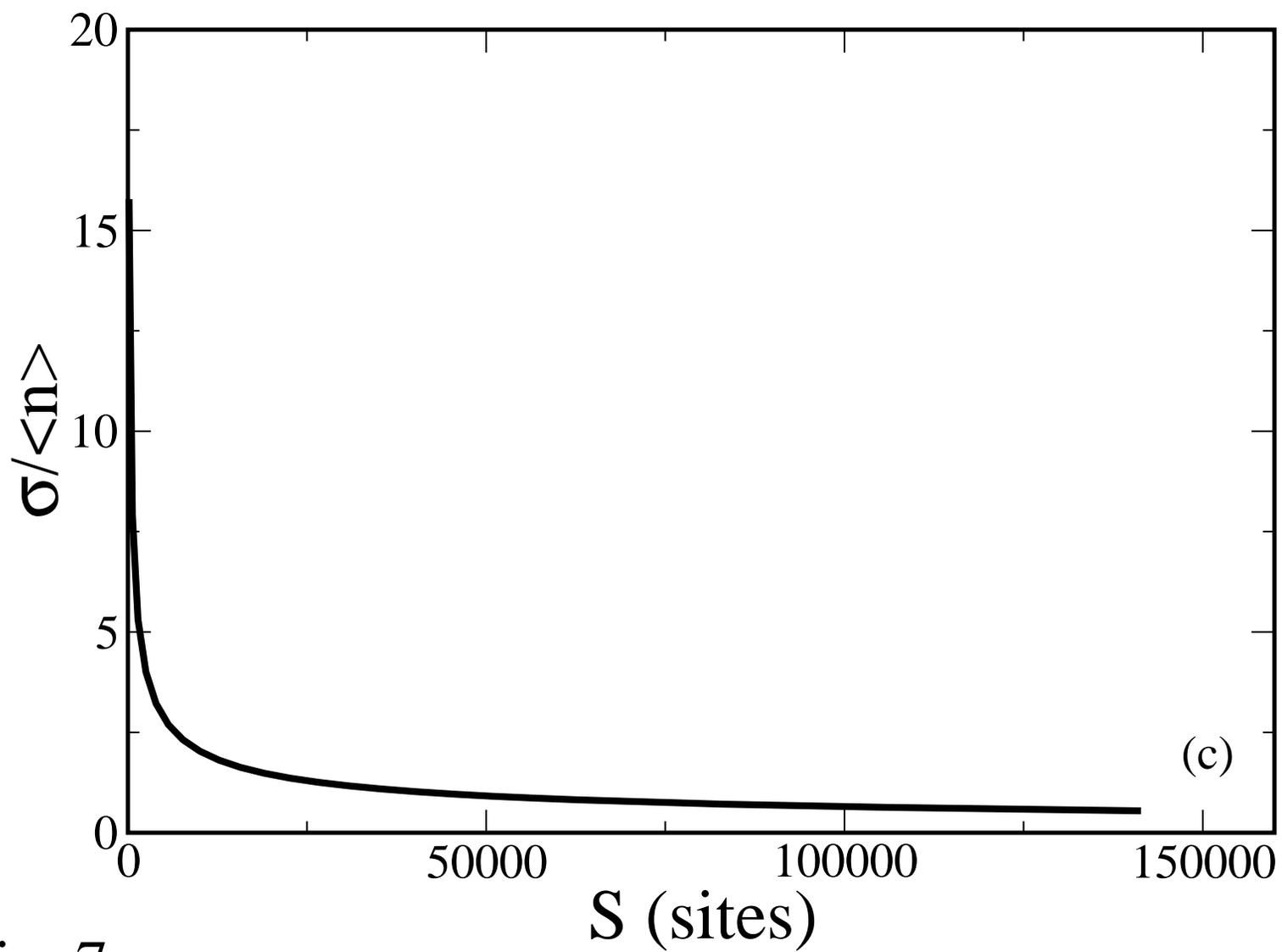
\end{document}